# A Machine Learning Approach for Design of Frequency Selective Surface based Radar Absorbing Material via Image Prediction

Vijay Kumar Sutrakar
*Scientist, ADE, DRDO*
Bangalore, India
vks.ade@gov.in

Anjana P K
*CE, ADE, DRDO*
Bangalore, India
pkanjana505@gmail.com

Sajal Kesharwani
*CE, ADE, DRDO*
Bangalore, India
kesharwanisajalvk18@gmail.com

Siddharth Bisariya
*CE, ADE, DRDO*
Bangalore, India
siddharthbisariya00@gmail.com

*Abstract*—The paper presents an innovative methodology for designing frequency selective surface (FSS) based radar absorbing materials using machine learning (ML) technique. In conventional electromagnetic design, unit cell dimensions of FSS are used as input and absorption coefficient is then predicted for a given design. In this paper, absorption coefficient is considered as input to ML model and image of FSS unit cell is predicted. Later, this image is used for generating the FSS unit cell parameters. Eleven different ML models are studied over a wide frequency band of 1GHz to 30GHz. Out of which six ML models (i.e. (a) Random Forest classification, (b) K- Neighbors Classification, (c) Grid search regression, (d) Random Forest regression, (e) Decision tree classification, and (f) Decision tree regression) show training accuracy more than 90%. The absorption coefficients with varying frequencies of these predicted images are subsequently evaluated using commercial electromagnetic solver. The performance of these ML models is encouraging and it can be used for accelerating design and optimization of high-performance FSS based radar absorbing material for advanced electromagnetic applications in future.

## I. INTRODUCTION

The design and optimization of electromagnetic materials for stealth and radar-absorbing structures (RAS) have undergone remarkable advancements, significantly influencing the development of modern stealth technology. Among these advancements, frequency selective surfaces (FSS) based radar- absorptive materials (RAM) has emerged as critical technologies. It offers unique capabilities in controlling electromagnetic (EM) wave propagation, crucial for reducing radar cross-sections (RCS) and enhancing stealth performance. Frequency Selective Surfaces are periodic structures that can filter electromagnetic waves for specific frequencies. As a result, they have been highly considered in applications such as radar stealth and communication systems. Indeed, by reflecting, transmitting, or absorbing waves due to frequency, FSS contributes to the design of stealth materials operating over wide frequency ranges [1]. Traditional mechanisms used in RAMs include dielectric loss, magnetic loss, or resistive loading, which, although effective, have associated penalties of increased weight and limited angular performance [2].

Recently, FSS based RAM has become increasingly recognized as break- through technologies. In contrast to the conventional RAM, FSS based RAM exploit their artificially designed electromagnetic properties, including tunable permittivity and permeability, to achieve phenomena such as wavefront manipulation, polarization control, and angle-independent absorption. The sub-wavelength, planar structure of these design avoids the disadvantages associated with traditional RAMs, such as bulkiness and high fabrication complexity, which is highly favorable for aerospace and stealth applications [3].

Current challenge in stealth aircraft design is to use ultra-thin and lightweight broadband radar-absorbing structures with polarization insensitivity and high angular performance. Moreover, the versatility of FSS based RAM for tuning reflection, transmission, and absorption properties enables the construction of advanced RAS tailored to specific operational scenarios [4].

Machine learning (ML) has further enhanced the design of FSS based RAM by providing efficient optimization of parameters and pattern recognition. The application of ML allows researchers to predict the electromagnetic responses of FSS RAM under various conditions of operation. This reduces computational efforts and improves the accuracy of stealth material design, thus marking a giant leap in stealth technology [5].

It will, therefore, create an enabling framework through the convergence of FSS RAM and machine learning for designing the next-generation stealth materials in the realm of electromagnetic wave manipulation as well as advanced radar evading techniques.

Directly determining FSS unit cell design without relying on highly computational intensive optimization procedures for a wide frequency band including angular performance, makes this approach highly advantageous. It may finally lead to significant saving in time and resources [6].

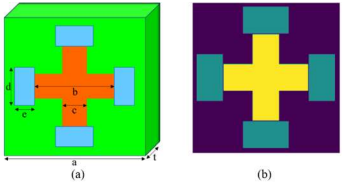

Fig. 1. Jerusalem cross metamaterial absorber (a) 3D view (b) 3D to 2D converted image

In this paper, the concept of implementing an inverse design methodology is used. Machine learning model is trained to generate the top-view image of a Jerusalem cross unit cell, given its target absorptive characteristics as input. This image prediction capability represents a significant advancement in the field of FSS based RAM design, as it leads to direct generation of unit cell designs tailored to specific performance criteria, without the need for extensive iterative simulations. The rest of the paper is as follows: Electromagnetic modeling and simulation details are provided in section II. Section III discusses the machine learning model. Results and discussions are presented in section IV, followed by concluding remarks in section V.

## II. ELECTROMAGNETIC MODELING AND SIMULATION DETAILS

In this study, the electromagnetic (EM) modeling and simulation of FSS-based radar absorbing materials have been carried out as a base step. A unit cell featuring a Jerusalem cross design was employed, where the cross is constructed from a perfectly electrically conducting (PEC) material (Orange in Fig 1. (a)). The cross is equipped with a fixed resistive value on its four arms (Blue in Fig 1. (a)). The overall structure of the unit cell includes three distinct layers. The first and third layers are composed of FR4 material, the second layer incorporates the Jerusalem cross design (refer [8] for further details). The total thickness t of the resulting sample is 2 mm. The reflection/absorption coefficient of the unit cell is determined using a Floquet port, alongside periodic boundary conditions (for the lateral surfaces) and a PEC boundary condition for the bottom layer. The analysis is carried out using a full-wave simulation technique in Ansys HFSS [7]. Extensive parametric sweeps are carried out, as depicted in Table 1. Subsequently, all the unit cells are saved as two-dimensional images. These images are used as output of the ML model. Further details are provided in the subsequent section.

TABLE I
UNIT CELL PARAMETERS AND THEIR RANGES

| Parameter | Values |
| --- | --- |
| b | 0.85, 1.15, 1.45, 1.75, 2.05, 2.35, 2.55 |
| c | 0.25, 0.55, 0.75 |
| d | 0.65, 0.95, 1.25, 1.55, 1.85, 1.95 |
| e | 0.25, 0.55, 0.75 |

## III. MACHINE LEARNING MODEL

In general, traditional modeling approaches that compute the electromagnetic response (i.e. absorption coefficient in the present case) of a predefined structure. In this work, an inverse design methodology is proposed for predicting the geometric configuration of the FSS-RAM (as an image) from its desired absorptive properties, as shown in Fig.2. The dataset (i.e. ∼ 460 samples) for this study is generated through comprehensive electromagnetic simulations [7]. Image of unit cell of each simulation is then saved along with its corresponding absorptivity over a range of frequencies varying from 1 GHz to 30 GHz (refer [8] for further details). The target images, initially in a dimension of (400, 400, 3), are preprocessed by converting them to 2-Dimensions (400, 400) with values 0, 1 and 2 for purple (glass fabric), green (resistive material) and yellow (conducting material) respectively, as shown in Fig. 1(b). Sample Ratio of 80:20 is considered for training and testing purpose. In the present setup, the absorptivity values are considered as the input features, while the corresponding unit cell images are the output of ML model. The following ML models are considered in the present study: (1) Random Forest classification (RFC) [9], (2) K-Neighbors Classification (KNC) [10], (3) Grid search regression (GSR) [11], (4) Random Forest regression (RFR) [12], (5) Decision tree classification (DTC) [13], (6) SVM regression (SVR) [14], (7) Linear regression (LIR) [15], (8) Ridge regression (RR) [16], (9) Decision tree regression (DTR) [17], (10) Elastic Net regression (ER) [18], and (11) Lasso regression (LR) [19]. The performance of the model is evaluated using R2 [20] and mean squared error (MSE).

## IV. RESULTS AND DISCUSSIONS

Initial study is carried out using Random Forest Regressor ML model [12]. Several hyperparameters of the Random Forest Regressor are fine-tuned. The overall R2 of 0.9122 for training and 0.6351 for testing is obtained. The performance of the Random Forest Regressor was evaluated using R2 and mean squared error (MSE). The true and predicted images for two initial test random samples are shown in Fig. 2.

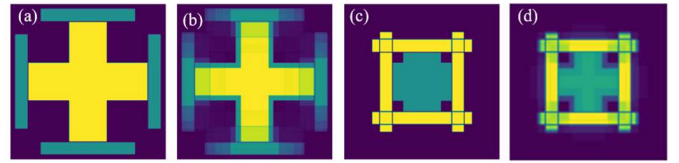

Fig. 2. True and predicted image of unit cell for (a) true image-1, (b) predicted image-1, (c) true image-2, and (d) predicted image-2 using Random Forest Regressor technique

Initially, the image values of 0, 1 and 2 during are considered during training for true images. However, after the prediction, certain modifications and assumptions were applied to the predicted unit cell images. These steps were necessary to align the predicted images with the practical requirements of the radar absorber. Firstly, a threshold algorithm is employed, where pixels with intensity values above a certain threshold were set to one material (e.g., PEC), and those below were set to another (e.g., dielectric substrate). This ensured clear delineation between different materials in the design. Three different cases are considered. In Case-1, values less than 0.4 is approximated as 0, values between 0.4 and 1.6 is approximated as 1, and any value above 1.6 is approximated as 2. In Case-2, values less than 0.5 is approximated as 0, values between 0.5 and 1.5 is approximated as 1, and any value above 1.5 is approximated as 2. In Case-3, values less than 0.6 is approximated as 0, values between 0.6 and 1.4 is approximated as 1, and any value above 1.4 is approximated as 2.

Initial assessment of true and predicted images for two initial test cases (refer Fig 2) are carried out. The corresponding absorption coefficient with varying frequencies are shown in Fig 3. However, in rest of the document, only Case-2 (i.e. values less than 0.5 is approximated as 0, values between 0.5 and 1.5 is approximated as 1, and any value above 1.5 is approximated as 2) is considered, as, it has closer match compared with the true value. The overall samples are split as 80% for training and 20% for testing. The details of train and test R-squared/accuracy as well as MSE for all the 11 models are shown in Table 2. Also, the execution time of each model is provided in Table 2. These experiments are carried out on a workstation with NVIDIA Quadro K620 graphics card and 64 GB RAM 500 GB ROM.

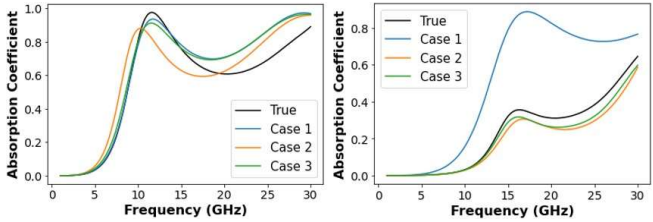

Fig. 3. Absorption coefficient of True and predicted images of Fig 2 for three different cases

True and predicted images for eight random datasets for all the 11 models are shown in Fig 4. Some of the regression models have predicted blurred images as shown in Fig 4. Those images are updated based on case-2 assumptions (i.e. values less than 0.5 is approximated as 0, values between 0.5 and 1.5 is approximated as 1, and any value above 1.5 is approximated as 2), as shown in Fig 5. Table 2, Fig 4 and 5 indicates that the RFC, KNC, GSR, RFR, DTC, and DTR based ML model predicts the FSS RAM image more accurately compared to remaining model. Images of Figures 4 and 5 are subsequently used for generating absorption coefficient using Ansys HFSS [7]. The absorption coefficient of true and predicted images (for six models which are performing better) with varying frequency from 1GHz to 30GHz is shown in Fig 6. A tolerance of ± 5% in unit cell parameter (and corresponding images) of true data is also used to predict the absorption coefficient, shown as shaded region in Fig 6. This will help to understand the absorption coefficient variations of predicted image compared to true image. The predicted absorption coefficients for all the models are encouraging. However, in few cases, KNC, GSR, and DTC models are not able to predict the absorption coefficients correctly across the band of frequencies. While the current study provides a solid foundation, several avenues for future research and development are envisioned for improving the current model. It includes expansion of our dataset by incorporating a wider range of unit cell with varying geometric parameters for improving model robustness and generalizability. This will enable improving its ability to predict configurations for a broader spectrum of applications. It is anticipated that by addressing these futuristic research directions, further advancements in the design and optimization of FSS based radar absorbing materials, will lead to more efficient and effective solutions for electromagnetic applications.

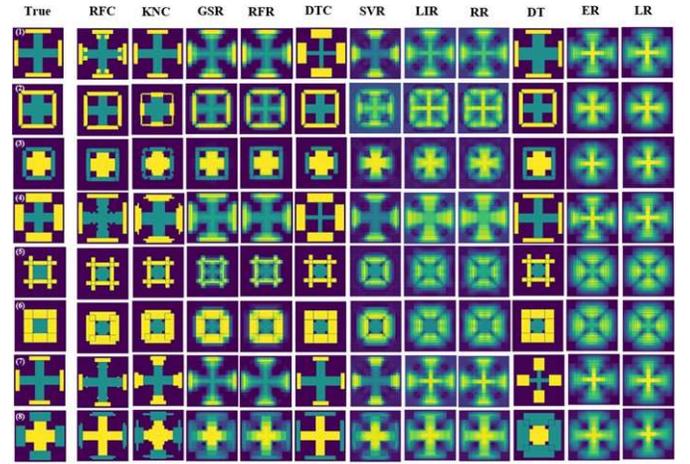

Fig. 4. True and predicted images of different ML models from test data

## V. CONCLUSIONS

In the present work, image based FSS RAM design is demonstrated using absorption coefficient as input to ML model. Eleven different ML models are studied over a wide frequency band of 1GHz to 30GHz. Out of which six ML models (i.e. (i) Random Forest classification, (ii) K- Neighbors Classification, (iii) Grid search regression, (iv) Random Forest regression, (v) Decision tree classification, and (vi) Decision tree regression) show training accuracy more than 90%. Successful prediction of the FSS unit cell images (geometric configurations of a Jerusalem cross unit cell) is demonstrated utilizing desired/targeted absorption coefficients over a wide frequency range. Results demonstrate the effectiveness of combining electromagnetic simulation with machine learning to facilitate rapid and efficient design of FSS based radar absorbing material. The ability to generate optimized unit cell images directly from targeted absorption coefficients represents a significant advancement in the field, offering a practical solution to the challenges of traditional iterative design methods.

TABLE II
PERFORMANCE OF DIFFERENT ML MODELS

| Sl. No. | Model | Train Accuracy | Test Accuracy | Train MSE | Test MSE | Execution Time (s) |
|---|---|---|---|---|---|---|
| 1 | Random forest classification (RFC) | 0.99 | 0.90 | 0.01 | 0.20 | 501 |
| 2 | K-Neighbors Classification (KNC) | 0.91 | 0.88 | 0.17 | 0.25 | 8 |
| 3 | Grid search regression (GSR) | 0.93 | 0.68 | 0.03 | 0.15 | 9154 |
| 4 | Random forest regression (RFR) | 0.91 | 0.64 | 0.04 | 0.16 | 202 |
| 5 | Decision tree classification (DTC) | 1.00 | 0.57 | 0.00 | 0.24 | 63 |
| 6 | SVM regression (SVR) | 0.62 | 0.55 | 0.18 | 0.20 | 877 |
| 7 | Linear regression (LIR) | 0.47 | 0.40 | 0.26 | 0.28 | 2 |
| 8 | Ridge regression (RR) | 0.47 | 0.40 | 0.26 | 0.28 | 0.48 |
| 9 | Decision tree regression (DTR) | 1.00 | 0.36 | 0.00 | 0.30 | 21 |
| 10 | Elastic Net regression (ER) | 0.32 | 0.30 | 0.33 | 0.33 | 49 |
| 11 | Lasso regression (LR) | 0.28 | 0.26 | 0.35 | 0.35 | 54 |

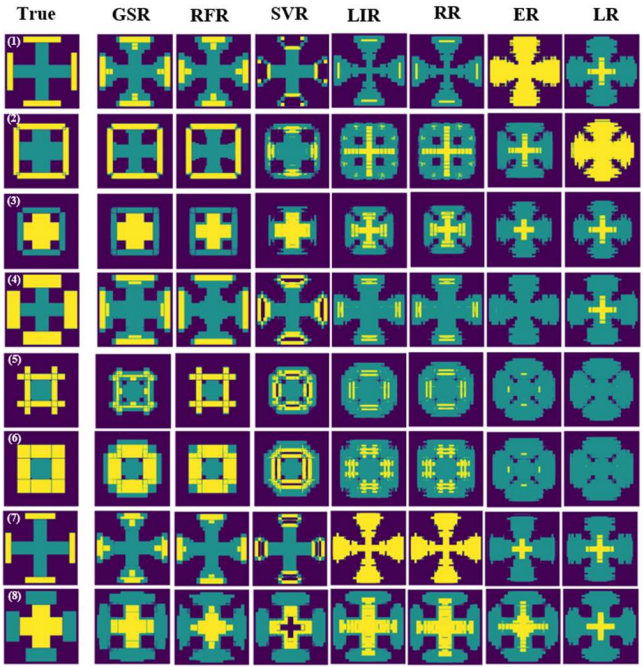

Fig. 5. Updated blurred images of seven regression models after dimensional adjustments as per Case 2

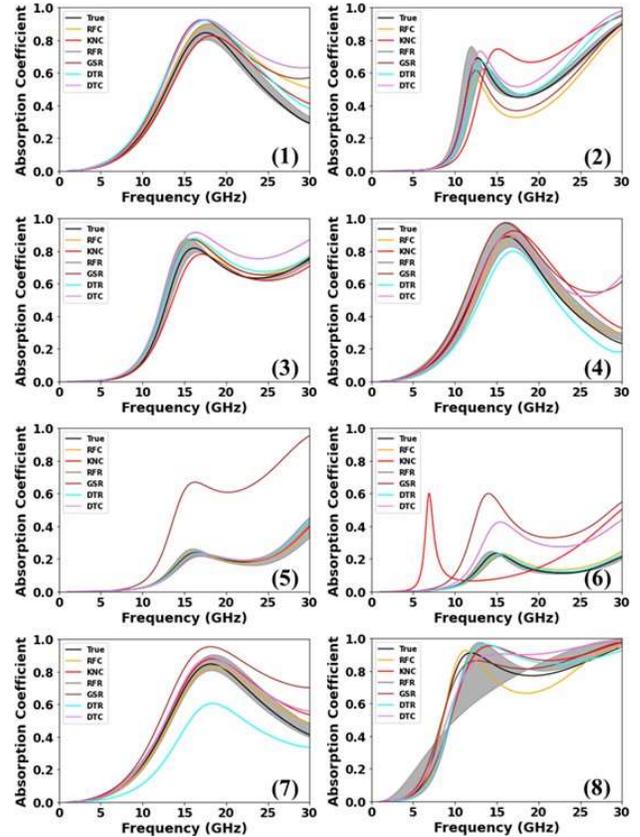

Fig. 6. Absorption coefficient for six different ML models for eight different test data samples

ACKNOWLEDGEMENT

Authors would like to thank Shri. Y Dilip, Director, Aeronautical Development Establishment, Mr. Manjunath S M, Technology Director and Mr. Diptiman Biswas, Group Director for their support during the research work carried out at ADE, DRDO.

REFERENCES

[1] Munk, B. A. Frequency Selective Surfaces: Theory and Design. Wiley-Interscience, 2000.
[2] Vinoy, K. J., & Jha, R. M. Radar Absorbing Materials: From Theory to Design and Characterization. Springer, 1996.
[3] Holloway, C. L., Kuester, E. F., Gordon, J. A., O'Hara, J., Booth, J., & Smith, D. R. (2012). "An overview of the theory and applications of metasurfaces: The two-dimensional equivalents of metamaterials." IEEE Antennas and Propagation Magazine, 54(2), 10-35.
[4] Cui, T. J., Qi, M. Q., Wan, X., Zhao, J., & Cheng, Q. (2014). "Coding metamaterials, digital metamaterials, and programmable metamaterials." Light: Science & Applications, 3(10), e218.
[5] Ma, W., Cheng, F., & Liu, Y. (2018). "Deep-learning-enabled on-demand design of chiral metamaterials." ACS Nano, 12(6), 6326-6334.
[6] Ghorbani, F., Beyraghi, S., Shabanpour, J., Oraizi, H., Soleimani, H., & Soleimani, M. (2021). Deep neural network-based automatic metasurface design with a wide frequency range. Scientific Reports, 11(1), 7102.
[7] ANSYS, "Ansys® HFSS Help, Release 22.1, Help System, Assigning Boundaries in HFSS," ANSYS, Inc., 2022.
[8] Anjana, P. K., Abhilash, P.V., Bisariya. S., and Sutrakar, V. K., "Inverse Approach for Metasurface Based Radar Absorbing Structure Design for Aerospace Applications Using Machine Learning Techniques," SAE Technical Paper 2024-26-0480, 2024, doi: 10.4271/2024-26-0480.
[9] Z. Bingzhen, Q. Xiaoming, Y. Hemeng and Z. Zhubo, "A Random Forest Classification Model for Transmission Line Image Processing," 2020 15th International Conference on Computer Science

& Education (ICCSE), Delft, Netherlands, 2020, pp. 613-617, doi: 10.1109/ICCSE49874.2020.9201900.
[10] Sabry, F. (2023). K Nearest Neighbor Algorithm: Fundamentals and Applications (Vol. 28). One Billion Knowledgeable.
[11] Raschka, S., Liu, Y. H., & Mirjalili, V. (2022). Machine Learning with PyTorch and Scikit-Learn: Develop machine learning and deep learning models with Python. Packt Publishing Ltd.
[12] Genuer, R., Poggi, J. M., Genuer, R., & Poggi, J. M. (2020). Random forests (pp. 33-55). Springer International Publishing.
[13] Suthaharan, S., & Suthaharan, S. (2016). Decision tree learning. Machine Learning Models and Algorithms for Big Data Classification: Thinking with Examples for Effective Learning, 237-269.
[14] Awad, M., Khanna, R., Awad, M., & Khanna, R. (2015). Support vector regression. Efficient learning machines: Theories, concepts, and applications for engineers and system designers, 67-80.
[15] Montgomery, D. C., Peck, E. A., & Vining, G. G. (2021). Introduction to linear regression analysis. John Wiley & Sons.
[16] Saleh, A. M. E., Arashi, M., & Kibria, B. G. (2019). Theory of ridge regression estimation with applications. John Wiley & Sons.
[17] Suthaharan, S., & Suthaharan, S. (2016). Decision tree learning. Machine Learning Models and Algorithms for Big Data Classification: Thinking with Examples for Effective Learning, 237-269.
[18] Chamlal, H., Benzmane, A., & Ouaderhman, T. (2024). Elastic net-based high dimensional data selection for regression. Expert Systems with Applications, 244, 122958.
[19] Aurelien Geron, Hands-on Machine Learning with Scikit-Learn, Keras, and TensorFlow, Second Edition, USA, y O'Reilly Media, 2019.
[20] Hou, J., Lin, H., Xu, W., Tian, Y. et al., "Customized Inverse Design of Metamaterial Absorber Based on Target-Driven Deep Learning Method," IEEE Access, 8:211849-211859, 2020.